\documentclass[aps,prl,preprint,tightenlines,superscriptaddress,showpacs,byrevtex]{revtex4}
\usepackage{amsmath}
\usepackage{amssymb}
\usepackage{graphicx}
\usepackage[dvips]{epsfig}

\usepackage[latin1]{inputenc}                                 
\usepackage{color}                                            
\usepackage{array}                                            
\usepackage{longtable}                                        
\usepackage{calc}                                             
\usepackage{multirow}                                         
\usepackage{hhline}                                           
\usepackage{ifthen}                                           

\usepackage{graphicx} 
\usepackage{dcolumn}  

\graphicspath{{ps}}

\begin{document}



\title{ \quad\\[0.5cm] Moments of the  photon energy spectrum from
  $B\to X_s \gamma$ decays
measured by Belle}


\affiliation{Aomori University, Aomori}
\affiliation{Budker Institute of Nuclear Physics, Novosibirsk}
\affiliation{Chiba University, Chiba}
\affiliation{Chonnam National University, Kwangju}
\affiliation{University of Cincinnati, Cincinnati, Ohio 45221}
\affiliation{University of Frankfurt, Frankfurt}
\affiliation{Gyeongsang National University, Chinju}
\affiliation{University of Hawaii, Honolulu, Hawaii 96822}
\affiliation{High Energy Accelerator Research Organization (KEK), Tsukuba}
\affiliation{Hiroshima Institute of Technology, Hiroshima}
\affiliation{Institute of High Energy Physics, Chinese Academy of Sciences, Beijing}
\affiliation{Institute of High Energy Physics, Vienna}
\affiliation{Institute for Theoretical and Experimental Physics, Moscow}
\affiliation{J. Stefan Institute, Ljubljana}
\affiliation{Kanagawa University, Yokohama}
\affiliation{Korea University, Seoul}
\affiliation{Kyoto University, Kyoto}
\affiliation{Kyungpook National University, Taegu}
\affiliation{Swiss Federal Institute of Technology of Lausanne, EPFL, Lausanne}
\affiliation{University of Ljubljana, Ljubljana}
\affiliation{University of Maribor, Maribor}
\affiliation{University of Melbourne, Victoria}
\affiliation{Nagoya University, Nagoya}
\affiliation{Nara Women's University, Nara}
\affiliation{National Central University, Chung-li}
\affiliation{National Kaohsiung Normal University, Kaohsiung}
\affiliation{National United University, Miao Li}
\affiliation{Department of Physics, National Taiwan University, Taipei}
\affiliation{H. Niewodniczanski Institute of Nuclear Physics, Krakow}
\affiliation{Nippon Dental University, Niigata}
\affiliation{Niigata University, Niigata}
\affiliation{Nova Gorica Polytechnic, Nova Gorica}
\affiliation{Osaka City University, Osaka}
\affiliation{Osaka University, Osaka}
\affiliation{Panjab University, Chandigarh}
\affiliation{Peking University, Beijing}
\affiliation{Princeton University, Princeton, New Jersey 08544}
\affiliation{RIKEN BNL Research Center, Upton, New York 11973}
\affiliation{Saga University, Saga}
\affiliation{University of Science and Technology of China, Hefei}
\affiliation{Seoul National University, Seoul}
\affiliation{Shinshu University, Nagano}
\affiliation{Sungkyunkwan University, Suwon}
\affiliation{University of Sydney, Sydney NSW}
\affiliation{Tata Institute of Fundamental Research, Bombay}
\affiliation{Toho University, Funabashi}
\affiliation{Tohoku Gakuin University, Tagajo}
\affiliation{Tohoku University, Sendai}
\affiliation{Department of Physics, University of Tokyo, Tokyo}
\affiliation{Tokyo Institute of Technology, Tokyo}
\affiliation{Tokyo Metropolitan University, Tokyo}
\affiliation{Tokyo University of Agriculture and Technology, Tokyo}
\affiliation{Toyama National College of Maritime Technology, Toyama}
\affiliation{University of Tsukuba, Tsukuba}
\affiliation{Utkal University, Bhubaneswer}
\affiliation{Virginia Polytechnic Institute and State University, Blacksburg, Virginia 24061}
\affiliation{Yonsei University, Seoul}
  \author{K.~Abe}\affiliation{High Energy Accelerator Research Organization (KEK), Tsukuba} 
  \author{K.~Abe}\affiliation{Tohoku Gakuin University, Tagajo} 
  \author{I.~Adachi}\affiliation{High Energy Accelerator Research Organization (KEK), Tsukuba} 
  \author{H.~Aihara}\affiliation{Department of Physics, University of Tokyo, Tokyo} 
  \author{K.~Aoki}\affiliation{Nagoya University, Nagoya} 
  \author{K.~Arinstein}\affiliation{Budker Institute of Nuclear Physics, Novosibirsk} 
  \author{Y.~Asano}\affiliation{University of Tsukuba, Tsukuba} 
  \author{T.~Aso}\affiliation{Toyama National College of Maritime Technology, Toyama} 
  \author{V.~Aulchenko}\affiliation{Budker Institute of Nuclear Physics, Novosibirsk} 
  \author{T.~Aushev}\affiliation{Institute for Theoretical and Experimental Physics, Moscow} 
% ...
\collaboration{The Belle Collaboration}

\noaffiliation

\begin{abstract}
We report preliminary measurements of the first (mean) and second
moment (variance) of the
inclusive photon energy spectrum in $B\to X_s\gamma$ decays, for
threshold values of the photon energy in the range 1.8--2.3~GeV as
measured in the
rest frame of the $B$-meson.
These results are obtained from
the Belle measurement of the spectrum, which used a data set
consisting of 152 million $B\overline{B}$ pairs collected
by the Belle detector at
the KEKB asymmetric-energy electron-positron collider operated on the
$\Upsilon(4S)$ resonance.
\end{abstract}


\maketitle

\tighten

{\renewcommand{\thefootnote}{\fnsymbol{footnote}}}
\setcounter{footnote}{0}
\section{Introduction}

The first (mean) and second (variance) moments of the
inclusive photon energy spectrum in $B\to X_s\gamma$ decays can be
used to determine the
Heavy Quark Effective Theory (HQET) parameters  $m_b$ and $\mu_\pi^2$
($\Lambda$ and
$\lambda_1$)~\cite{Ligeti:1999ea,Bauer:1997fe,Bigi:2002qq,Benson:2004sg}.
These parameters play a crucial role in
the extraction of the CKM matrix elements $|V_{cb}|$ and
$|V_{ub}|$~\cite{Bauer:2002sh,Mahmood:2002tt,Aubert:2004aw,Bauer:2004ve,Lange:2005yw}.
Furthermore, the measurements can be compared with predictions made
using different theoretical
treatments~\cite{Bauer:1997fe,Ligeti:1999ea,Benson:2004sg,Gardi:2005mf,Neubert:2005nt}.

Belle reported moment measurements using an energy threshold of
1.8~GeV in Ref~\cite{Koppenburg:2004fz}; in this note we report on a 
further analysis of the measured spectrum yielding moment measurements
at photon energy thresholds in the range 1.8--2.3 GeV as measured
in the rest frame of the $B$-meson. Correlation
coefficients between the measurements are also given; these are
necessary for inclusion in an analysis
to determine HQET parameters~\cite{Bauer:2004ve}.

\section{Review of the Belle Measurement}
The $B\rightarrow X_s \gamma$
photon energy spectrum measurement at Belle is briefly reviewed.
The analysis used data samples amounting to
$140\,\mathrm{fb}^{-1}$ and $15\,\mathrm{fb}^{-1}$ of integrated
luminosity taken at (ON) and $60\,\mathrm{M}\mathrm{eV}$ below (OFF) the $\Upsilon(4S)$ resonance energy
respectively. 

The analysis procedure involved
reconstructing photon candidates with energy greater than
$1.5\,\mathrm{G}\mathrm{eV}$ as
measured in the $\Upsilon(4S)$ rest frame. Photon candidates were vetoed if they had
a high likelihood of originating from $\pi^0$ or $\eta$ decays to two
	photons. The likelihood, modelled in Monte Carlo (MC),
was calculated as a function of the combined invariant mass
of the photon candidate paired with another photon reconstructed in the
event, and the energy and polar angle of that
other photon in the laboratory frame. 

In general, the background of photons from the $e^+e^-\rightarrow
q\overline{q}$ continuum dominates. It
is suppressed through use of event shape variables, which
are used as the inputs to two Fisher discriminants~\cite{Fisher}.
The first discriminant is used to
distinguish spherically-shaped $B\overline{B}$ events from jet-like
continuum events and includes the
Fox-Wolfram moments~\cite{Fox}, the thrust calculated using all particles
detected in the event including and excluding the
candidate photon, and the angles of the corresponding thrust axes with
respect to the beam and candidate photon directions, respectively. The second
discriminant is
designed to exploit the topology of $B\rightarrow X_s\gamma$ events by 
utilising the energy sum of detected particles measured in three        
angular regions,
$\alpha^* < 30^\circ$,
$30^\circ\le\alpha^\ast\le140^\circ$, and
$\alpha^\ast>140^\circ$, where $\alpha^\ast$ 
is the angle to the candidate photon.
After cuts the remaining continuum background is removed
by subtracting scaled OFF data yields from those of ON data. The
scaling factor is obtained by taking the ON-to-OFF ratio of
measured luminosities corrected for the difference in cross section
between ON and OFF resonance centre of mass energies. 

Backgrounds from $B$ decays are estimated from 
MC and scaled according to studies using data wherever possible and
then subtracted from the data.
Their contributions include:
\begin{itemize}
\item photons from $\pi^0$ and $\eta$ (veto leakage);
\item other real photons, mainly from  $\omega$, $\eta'$, and $J/\psi$;
\item clusters in the calorimeter not due to single photons (mainly electrons 
interacting with matter, $K^0_L$ and $\bar{n}$);
\item beam background.
\end{itemize}
  
\begin{figure}[htbp!]
  \begin{center}
    \begin{tabular}{cc}
      \includegraphics[width=0.50\columnwidth]{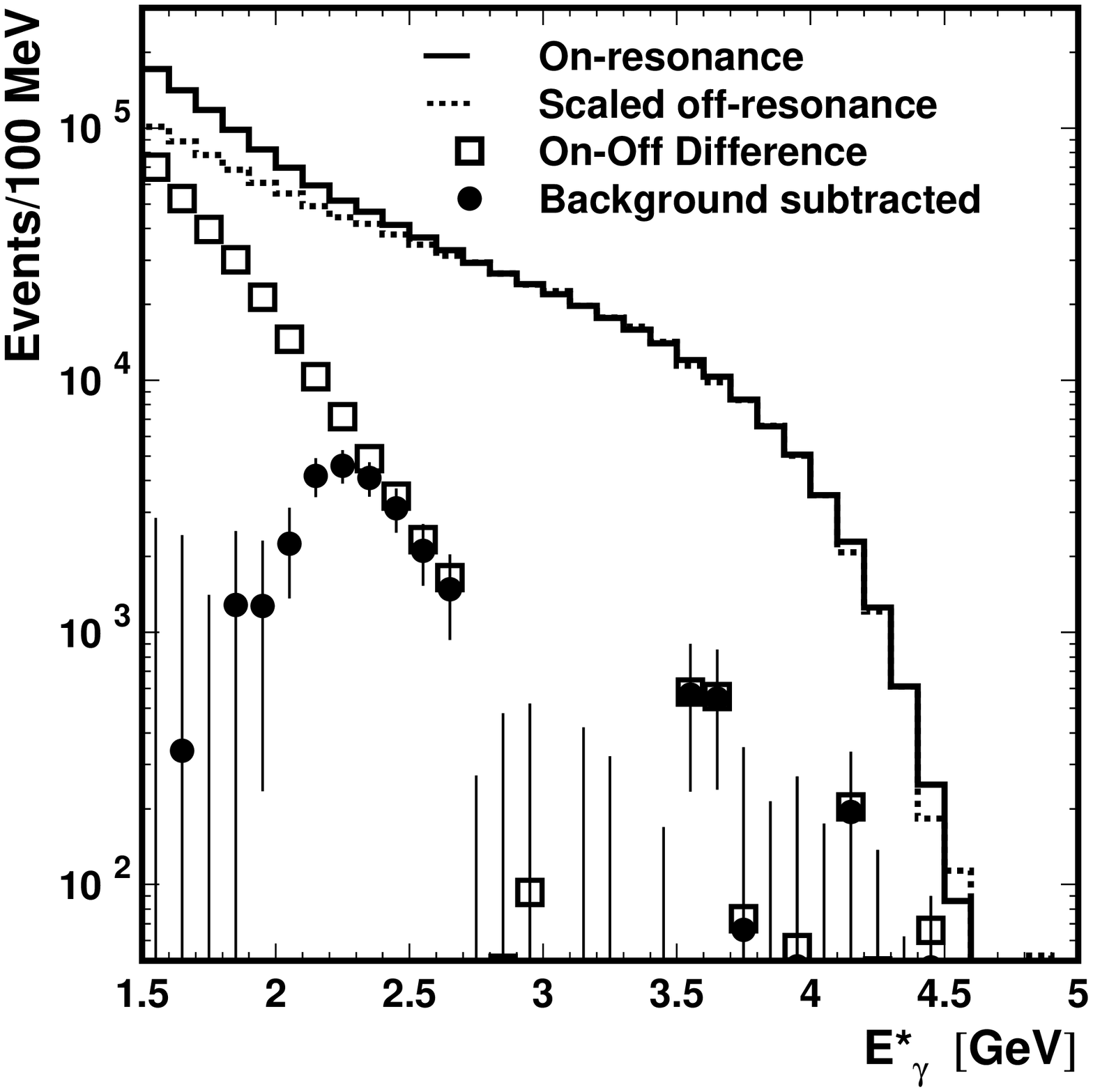} &
      \includegraphics[width=0.50\columnwidth]{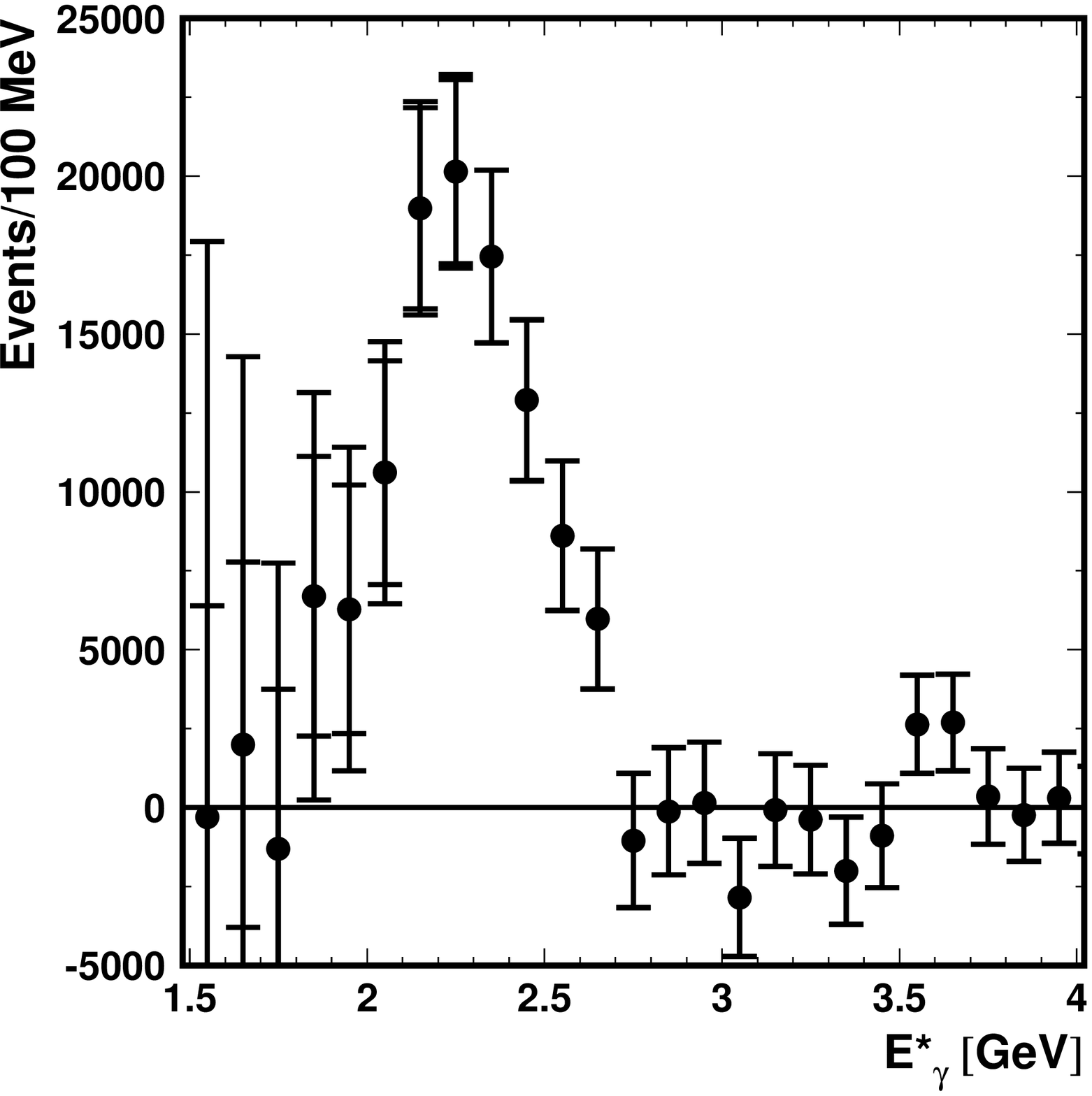}\\
      (a) & (b) \\
    \end{tabular}
 \end{center}
\caption{From~\cite{Koppenburg:2004fz}.
  (a) Photon energy spectra in the $\Upsilon(4S)$ frame.
  (b) Efficiency-corrected photon energy spectrum.
  The two error bars show the statistical and total errors.
         \label{fig:bsgspec}}
\end{figure}

The photon spectra for ON and scaled OFF data samples along with the
results of subsequent background subtractions are plotted in
Fig.~\ref{fig:bsgspec}(a). The $B\rightarrow X_s \gamma$ photon
energy spectrum that has been corrected for efficiency
is shown in Fig.~\ref{fig:bsgspec}(b).
The analysis measured the branching fraction,
\begin{equation}
  \mathcal{B}(B\rightarrow X_s \gamma) = (3.55 \pm
  0.32^{+0.30+0.11}_{-0.31-0.07}) \times 10^{-4},
\end{equation}
where the errors are statistical, systematic and theoretical,
respectively. This result agreed with the latest theoretical 
calculations~\cite{Hurth:2003dk,Gambino:2001ew}, as well as with
previous measurements made by CLEO~\cite{CLEOb2g} and
Belle~\cite{Belleb2g}.

\section{Moment measurements}
We follow the procedure used in the published analysis~\cite{Koppenburg:2004fz},
with some slight variations.
No attempt is made to correct for the part of
the spectrum that is not measured with a satisfactory precision
\emph{i.e} from energy below 1.8 GeV. 
We apply lower energy threshold
cuts as measured in the $\Upsilon(4S)$ rest 
frame ($E^*_\mathrm{cut}$) to the efficiency corrected spectrum, from which we obtain truncated first
and second moments. Corrections are applied to recover the
moments such that the lower
energy thresholds correspond to quantities measured
in the $B$-meson rest frame ($E_\mathrm{cut}$).   

A simple procedure is used to unfold the effects
of detector resolution, the small $B$-meson boost in the
$\Upsilon(4S)$ frame, and that of the 100 MeV wide bins.
We define the first moment as  $\left< E_\gamma \right>$  (mean)
and the second moment as $\Delta E^2_\gamma \equiv
 \left < E^2_\gamma \right>-\left< E_\gamma \right>^2$ (variance). 
The corrections are as follows:
\begin{description}
\item[$\mathbf{B}$-meson boost:]
The small momentum $|\vec{p}_B|$ of the $B$ meson in the $\Upsilon(4S)$
frame reduces the mean of the energy distribution on average by
\[
    \frac{\Delta E_{\mathrm{boost}}}{\left< E \right>} = 1 - \frac{2m_B}{m_{\Upsilon(4S)}}\approx 0.002 
    \]
    and adds a Doppler broadening of
    \[
    \left( \frac{\left<E^*_\gamma\right>
      |\vec{p}_B|}{\sqrt{3}m_B}
    \right)^2\approx 0.006\,\mathrm{GeV^2}
    \]
    to the second moment.
\item[Binning:]
    Using $100\,\mathrm{MeV}$ bins artificially adds a contribution
    \[
    \left(
    \frac{1}{\sqrt{12}}0.1\,\mathrm{GeV}
    \right)^2 \approx 0.0008\,\mathrm{GeV^2},
    \]
    to the second moment.

  \item[Energy Resolution:] The energy measurement resolution has the
    effect of broadening the spectrum, adding
    \[
    W^2_{\mathrm{ECL}} = \left( \delta_{\mathrm{ECL}}
    \left<E^*_\gamma \right> \right)^2 \approx 0.004\,\mathrm{GeV^2}
    \] to the second moment, where $\delta_{\mathrm{ECL}}=2.8\%$ is the
    energy resolution.
  \item[Bias correction:] The previous corrections can be classified as
  model independent and do not together compensate
  for the full effect of measurement. An additional bias correction
  derived from the signal MC sample is implemented. This sample, which
  was also used to correct the measured spectrum for acceptance,
  was generated as a weighted sum of $B\to K^\ast\gamma$ decays,
  where $K^\ast$ is any known spin-$1$ resonance
  with strangeness $S=1$ where the relative weights are obtained by matching 
  the total photon spectrum to a theoretical model~\cite{KN}.
  The bias correction is calculated as the difference of the true moment and
  the moment measured from the signal MC sample once all aforementioned
  corrections have been applied. The low-energy tail of the photon energy
  resolution, in general, decreases the first moment but we found the
  correction to be model dependent and therefore preferred to classify
  it as a bias correction. Bias corrections for both
  moments are shown in Table~\ref{tab:bias}.
      
  \begin{table}
    \caption{Bias corrections as a function of $E_\mathrm{cut}$ for
      the first and second moments.\label{tab:bias}}
    \begin{tabular}{|c|c|c|}
      \hline\hline
       & \multicolumn{2}{c|}{Bias} \\
      $E^*_{\mathrm{cut}}$ (GeV)& $\left< E_\gamma \right>$  (\%) & $\Delta E^2_\gamma$  (\%) \\
      \hline
      1.8 & +2.0   &  0.0\% \\
      1.9 & +1.6   & -0.4\% \\
      2.0 & +1.2   & -7.1\% \\
      2.1 & +0.8   & -17.4\% \\
      2.2 & +0.2   & -35.3\% \\
      2.3 & -0.3   & -57.9\% \\
      \hline
    \end{tabular}
  \end{table}. 
\end{description}

\begin{figure}[htbp!]
  \begin{center}
    \begin{tabular}{cc}
      \includegraphics[width=0.50\columnwidth]{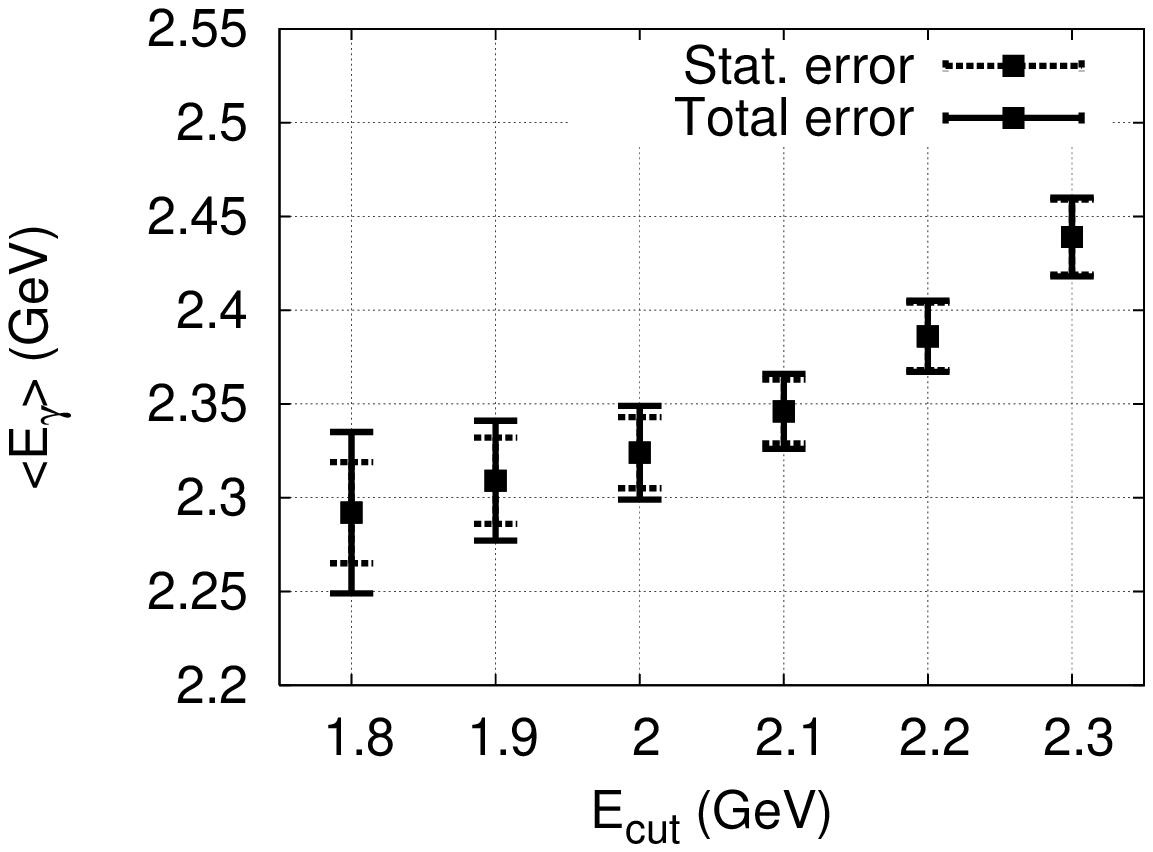} &
      \includegraphics[width=0.50\columnwidth]{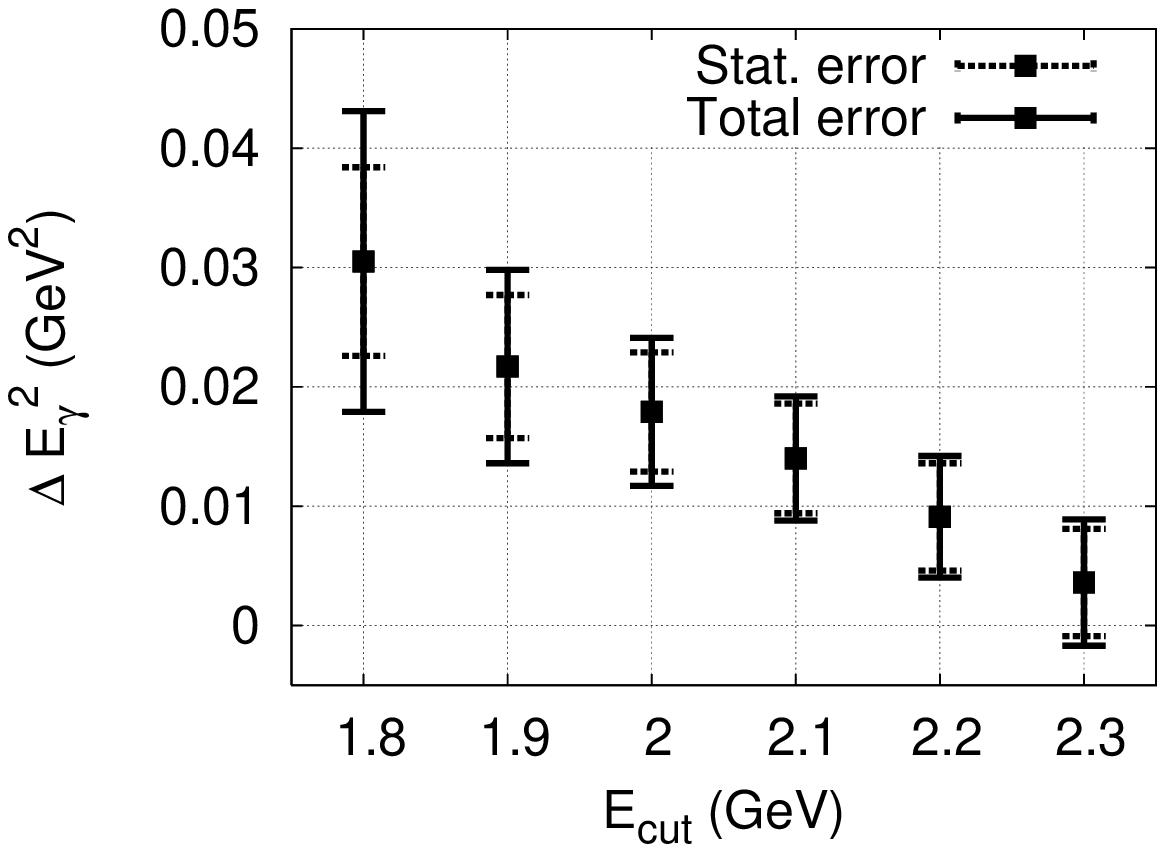}\\
      (a) & (b) \\
    \end{tabular}
 \end{center}
\caption{
  $B\rightarrow X_s \gamma$ photon spectrum  (a) mean and
  (b) variance as a function of $E_{\mathrm{cut}}$.
         \label{fig:moments1}}
\end{figure}

A systematic uncertainty on the moments stems from the
systematic error on the binned signal yields and is referred to as the
``scaling'' systematic uncertainty. Systematic variations
that were considered for the branching fraction measurement are
implemented, namely we: vary the number of $B\overline{B}$ events simultaneously
with the ON to OFF data ratio; vary within uncertainty
the ON and OFF selection efficiency
  difference; vary the fitting functions that correct for DATA
  and MC differences in response to the $B\overline{B}$ background; 
  vary by $\pm 20\%$ background from $\eta^\prime$, $\omega$ and
  bremsstrahlung; vary the $\eta$ veto efficiency for real $\eta$ mesons;
  use an ``alternate'' signal MC that favours high-mass resonances
  decaying into high-multiplicity final states and is the same sample
  that was used to estimate the model dependence in the branching
  fraction measurement; and vary the photon detection efficiency by $\pm
  2.3\%$ for both signal and backgrounds. We also implement
  a $\pm 50\%$ variation on the bias correction for the first moment
  while for the
  second moment the correction is re-calculated using the alternate
  signal MC sample. In addition for the second moment a variation, that neglects the lower
energy tail in the resolution and assumes a Gaussian
model, where $\delta_{\mathrm{ECL}}=1.9\%$, is implemented. The observed
difference due to each variation listed above is
assigned as a systematic uncertainty. We also assign a $\pm 100\%$
uncertainty on
the binning correction for the second moment. The net systematic error
is calculated from the sum in quadrature of the individual
uncertainties. Table~\ref{tab:moments} lists the moment measurements
and they are plotted in Figure~\ref{fig:moments1}.
Table~\ref{tab:errors} gives a full account of the uncertainties
for both the first and second moment with varying values of the lower energy
threshold $E_{\mathrm{cut}}$=1.8--2.3 GeV.

The statistical uncertainty and scaling
systematic uncertainty have been calculated using a toy MC study.
We generated numerous random spectra according to the
measured spectrum with the bin yields and their uncertainties
corresponding to the mean and standard deviation of
a Gaussian random variable, respectively. The moments and their
fluctuations with respect to each other were measured for each
generated spectrum, and finally averaged to yield the covariance
matrix, from which the uncertainties due to statistics and systematics scaling
were obtained. The covariance matrix was also obtained from systematic
variations due to the corrections to the moments. Table~\ref{tab:corr} shows
the correlation coefficients calculated from the combined
covariance matrix. 

At $E_{\mathrm{cut}}=1.8\,\mathrm{GeV}$ the uncertainty on the first moment is dominated by the systematic
uncertainty from scaling and the bias
correction. The second moment systematic uncertainty
is dominated by the bias correction. In total the systematic error is
larger than the statistical error. This circumstance is reversed at
$E_{\mathrm{cut}}=2.3\,\mathrm{GeV}$ where the statistical error
dominates. The amount of OFF data used in the analysis
limits the statistical precision. For increasing $E_{\mathrm{cut}}$
the systematic uncertainty reduces due to decreasing $B\overline{B}$ event
backgrounds.
Note the statistical and systematic
errors on the moments at $E_{\mathrm{cut}}=1.8\,\mathrm{GeV}$ are
slightly different to the values quoted in our
publication~\cite{Koppenburg:2004fz}. This is
due to both the use of toy MC as well as the bias correction
uncertainty; the latter has increased the systematic uncertainty in
the second moment. 

\begin{figure}[htbp!]
  \begin{center}
    \begin{tabular}{cc}
      \includegraphics[width=0.50\columnwidth]{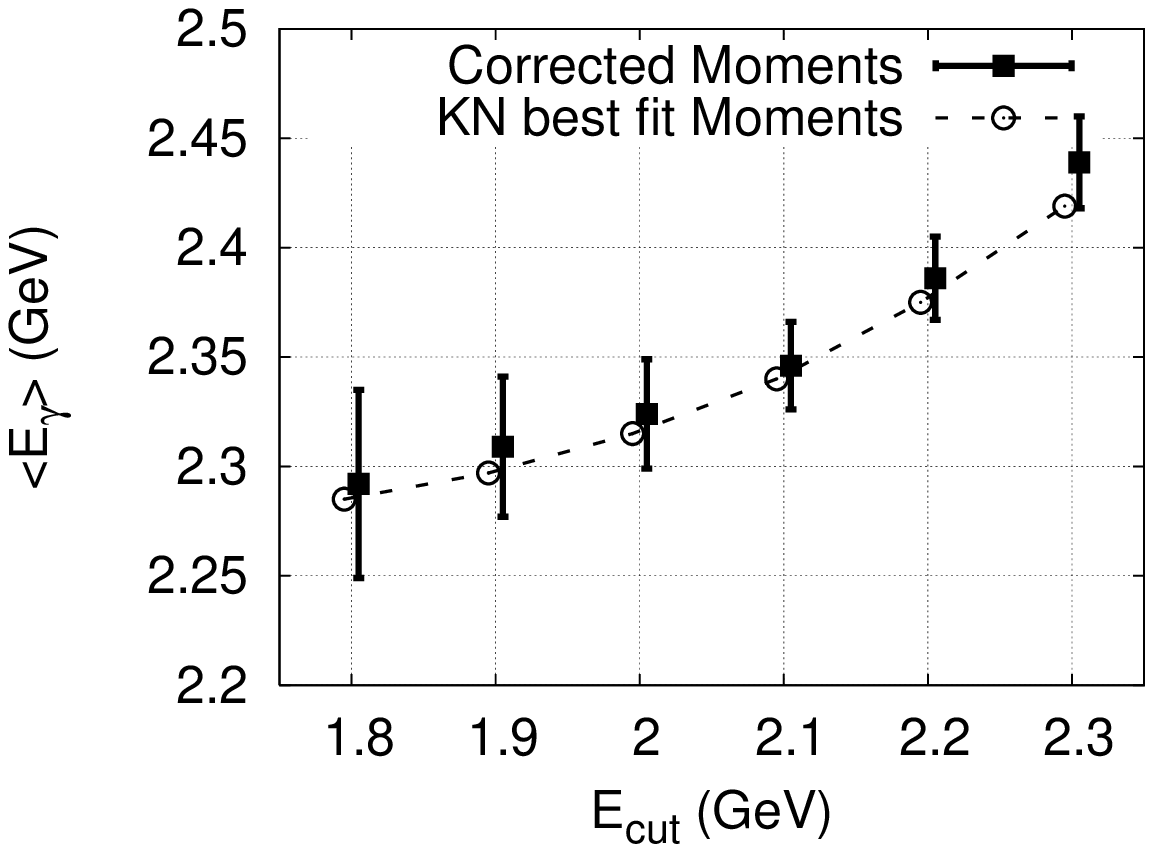} &
      \includegraphics[width=0.50\columnwidth]{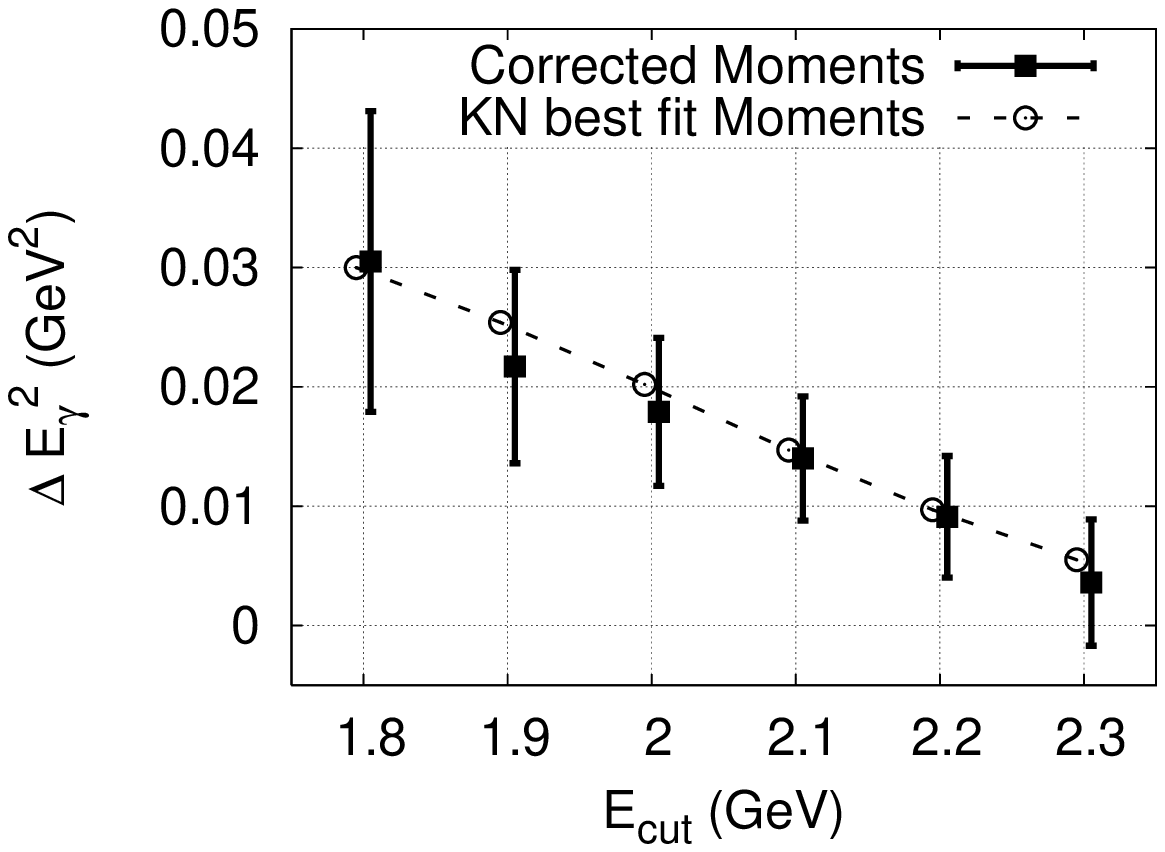}\\
      (a) & (b) \\
    \end{tabular}
 \end{center}
\caption{ Comparison of
  $B\rightarrow X_s \gamma$ photon spectra  (a) mean and
  (b) variance as a function of $E_{\mathrm{cut}}$ between our
  measured moments and those predicted in the Kagan-Neubert
  prescription using corresponding values of the parameters that best
  fit the spectrum.
         \label{fig:moments}}
\end{figure}

We performed a cross check where we measured the moments using a
method that utilises the analysis that determined parameters in the Kagan-Neubert
prescription (KN)~\cite{KN}, which were found to best fit
our
spectrum~\cite{Limosani:2004jk}
($m_b(\mathrm{KN})=4.62\,\mathrm{GeV}/c^2,{\mu_\pi^2}(\mathrm{KN})=0.40\,\mathrm{GeV^2}/c^2$).
We generated the photon spectrum
in the rest frame of the $B$-meson with these parameters as input and
extracted the moments for $E_\mathrm{cut}$=1.8--2.3 GeV. The results are plotted
in Fig.~\ref{fig:moments} along with the measured moments. We find
very good agreement between the moments measured from these
independent methods for all but the first moment at $E_{\mathrm{cut}}=2.3$
GeV. Even for that case the results still agree. 
The moment measurements agree within uncertainty
with the CLEO result~\cite{CLEOb2g} as well with preliminary
measurements reported by BaBar~\cite{Mommsen,Bucci:2005cb}.

\section{Summary}
We have reported preliminary measurements of the first (mean)
and second (variance) moment and their correlations, of
the $B\rightarrow X_s \gamma$ photon energy spectrum measured by Belle,
for lower threshold values of the photon energy in the range
1.8--2.3~GeV, as measured in the $B$-meson rest frame. These can be
used to determine HQET parameters $m_b$ and $\mu_\pi^2$ in the kinetic
scheme or equivalently $\bar{\Lambda}$ and $\lambda_1$ in the $1S$
scheme. 

\begin{table}
  \caption{Moment measurements depending on the lower energy threshold
  $E_{\mathrm{cut}}\,(\mathrm{GeV})$, where the first error is
  statistical and the second is systematic.\label{tab:moments}}
  \begin{tabular}{|c|ccc|ccc|}
    \hline\hline
    $E_{\mathrm{cut}}\,(\mathrm{GeV})$ & & $\left < E_\gamma \right >\,(\mathrm{GeV})$
    & & & $\Delta E^2_\gamma\,(\mathrm{GeV}^2)$ & \\\hline
    1.8 &  & $2.292 \pm   0.027 \pm   0.033 $ & &  & $0.0305 \pm
    0.0079 \pm  0.0099 $ & \\
    1.9 &  & $2.309 \pm   0.023 \pm   0.023 $ & &  & $0.0217 \pm
    0.0060 \pm  0.0055 $ & \\
    2.0 &  & $2.324 \pm   0.019 \pm   0.016 $ & &  & $0.0179 \pm
    0.0050 \pm  0.0036 $ & \\
    2.1 &  & $2.346 \pm   0.017 \pm   0.010 $ & &  & $0.0140 \pm
    0.0046 \pm  0.0024 $ & \\
    2.2 &  & $2.386 \pm   0.018 \pm   0.005 $ & &  & $0.0091 \pm
    0.0045 \pm  0.0025 $ & \\
    2.3 &  & $2.439 \pm   0.020 \pm   0.004 $ & &  & $0.0036 \pm
    0.0045 \pm  0.0028 $ & \\\hline
    \hline
  \end{tabular}
\end{table}

\begin{table}
\caption{Uncertainties contributing to the moment
  measurements depending on the lower energy threshold
	$E_{\mathrm{cut}}\,(\mathrm{GeV})$. \label{tab:errors}}
\begin{tabular}{|l||c|c|c|c|c|c||c|c|c|c|c|c|}
  \hline
  \multicolumn{1}{|c||}{Source of} &
  \multicolumn{6}{c||}{$\delta(\left < E_\gamma \right >) \,(\mathrm{GeV})$} &
  \multicolumn{6}{c|}{$\delta(\Delta E^2_\gamma)\,(\mathrm{GeV}^2)$} \\
   \multicolumn{1}{|c||}{systematic} & 1.8 & 1.9 & 2.0 & 2.1 & 2.2 & 2.3 & 1.8 & 1.9 & 2.0 & 2.1 & 2.2 & 2.3 \\
    \hline
    Scaling & 0.021 & 0.012 & 0.006 & 0.003 & 0.002 & 0.001
    & 0.0060 & 0.0027 & 0.0009 & 0.0003 & 0.0001 & 0.0001 \\
    \hline						 
    Energy resolution &  & & & & &
    & 0.0020 & 0.0020 & 0.0021 & 0.0022 & 0.0023 & 0.0024 \\
    Binning & & & & & &
    & 0.0008 & 0.0008 & 0.0008 & 0.0008 & 0.0008 & 0.0008 \\
    Bias      & 0.022 & 0.018 & 0.014 & 0.009 & 0.002 & 0.003

    & 0.0068 & 0.0040 & 0.0026 & 0.0005 & 0.0004 & 0.0011 \\ 
    \hline						 
    No. of $B\overline{B}$ & 0.004 & 0.001 & 0.001 & 0.002 & 0.002 & 0.002
    & 0.0018 & 0.0010 & 0.0006 & 0.0005 & 0.0005 & 0.0004 \\ 
    ON/OFF eff. & 0.000  & 0.000  & 0.000  & 0.000  & 0.000 & 0.000
    & 0.0000 & 0.0000 & 0.0000 & 0.0000 & 0.0000 & 0.0000 \\ 
    MC/DATA eff. & 0.005 & 0.003 & 0.002 & 0.001 &
    0.000 & 0.000 & 0.0010 & 0.0004 & 0.0001 & 0.0000 & 0.0000 & 0.0000
  \\  
    Other $B\overline{B}$ $\gamma$ & 0.010 & 0.004 & 0.002 & 0.000 & 0.000 & 0.000
    & 0.0024 & 0.0008 & 0.0002 & 0.0000 & 0.0000 & 0.0000 \\
    $\eta$ veto on $\eta$  & 0.001 & 0.001 & 0.000 & 0.000 & 0.000 & 0.000
    & 0.0003 & 0.0001 & 0.0000 & 0.0000 & 0.0000 & 0.0000 \\
    Signal MC & 0.004 & 0.004 & 0.004 & 0.004 & 0.004 & 0.003
    & 0.0007 & 0.0005 & 0.0004 & 0.0003 & 0.0002 & 0.0000 \\ 
    $\gamma$ detection eff. & 0.001 & 0.001 & 0.000 & 0.000 & 0.000 & 0.000
    & 0.0003 & 0.0001 & 0.0000 & 0.0000 & 0.0000 & 0.0000 \\
    \hline						 
    Total systematic & 0.033 & 0.023 & 0.016 & 0.010 & 0.005 & 0.004
    & 0.0099 & 0.0055 & 0.0036 & 0.0024 & 0.0025 & 0.0028 \\ \hline\hline						 
    Statistical error & 0.027 & 0.023 & 0.019 & 0.017 & 0.018 &  0.020 
    & 0.0079 & 0.0060 & 0.0050 & 0.0046 & 0.0045 & 0.0045 \\ \hline\hline
    Total error & 0.043 & 0.032 & 0.025 & 0.020 & 0.019 &  0.021 
    & 0.0126 & 0.0081 & 0.0062 & 0.0052 & 0.0051 & 0.0053 \\   
    \hline
\end{tabular}
\end{table}  

\begin{table}
  \caption{Correlation coefficients between the moment measurements.
    The calculation takes into account both statistical and systematic
    uncertainties.\label{tab:corr}}
  \begin{center}
    \begin{tabular}{|cc|cccccc|cccccc|} \hline\hline
      \multicolumn{2}{|c|}{$E_{\mathrm{cut}}$}     &
      \multicolumn{6}{c|}{$\left< E_\gamma \right>$}  &
      \multicolumn{6}{c|}{$\Delta E^2_\gamma$ } 
      \\ 
      \multicolumn{2}{|c|}{($\mathrm{GeV}$)} &
      1.8  & 
      1.9  &
      2.0  &
      2.1  &
      2.2  &
      2.3  &
      1.8  &
      1.9  &
      2.0  &
      2.1  &
      2.2  &
      2.3  
      \\ \hline
      \multirow{6}{8mm}{$\left< E_\gamma \right>$} & 
      1.8 & 
      $\phantom{-}1.00$ &
      $\phantom{-}0.79$	&
      $\phantom{-}0.68$	&
      $\phantom{-}0.56$	&
      $\phantom{-}0.38$	&
      $\phantom{-}0.22$	&
      $-0.46$	&
      $-0.18$	&
      $-0.01$	&
      $\phantom{-}0.04$	&
      $\phantom{-}0.01$	&
      $-0.01$
\\
&
1.9 &
&
$\phantom{-}1.00$ &
$\phantom{-}0.82$	&
$\phantom{-}0.70$	&
$\phantom{-}0.52$	&
$\phantom{-}0.33$	&
$-0.06$	&
$-0.21$	&
$\phantom{-}0.05$	&
$\phantom{-}0.12$	&
$\phantom{-}0.10$	&
$\phantom{-}0.07$	
\\
&
2.0 &
&
&
$\phantom{-}1.00$ &
$\phantom{-}0.86$	&
$\phantom{-}0.67$	&
$\phantom{-}0.47$	&
$-0.14$	&
$\phantom{-}0.15$	&
$\phantom{-}0.12$	&
$\phantom{-}0.23$	&
$\phantom{-}0.20$	&
$\phantom{-}0.17$	
\\		
&
2.1 &
&
&
&
$\phantom{-}1.00$ &
$\phantom{-}0.84$	&
$\phantom{-}0.65$	&
$\phantom{-}0.27$	&
$\phantom{-}0.37$	&
$\phantom{-}0.43$	&
$\phantom{-}0.42$	&
$\phantom{-}0.39$	&
$\phantom{-}0.34$	
\\		
&
2.2	 
&
&
&
&
&
$\phantom{-}1.00$ &
$\phantom{-}0.86$	&
$\phantom{-}0.38$	&
$\phantom{-}0.55$	&
$\phantom{-}0.67$	&
$\phantom{-}0.75$	&
$\phantom{-}0.66$	&
$\phantom{-}0.61$	
\\		
&
2.3	
&	 
& 
&
&
&
&
$\phantom{-}1.00$ &
$\phantom{-}0.43$	&
$\phantom{-}0.63$	&
$\phantom{-}0.79$	&
$\phantom{-}0.91$	&
$\phantom{-}0.88$	&
$\phantom{-}0.79$	
\\		\hline
\multirow{6}{8mm}{$ \Delta E^2_\gamma $} &
1.8 &	
&
&
&
&
&
&
$\phantom{-}1.00$ &
$\phantom{-}0.72$	&
$\phantom{-}0.63$	&
$\phantom{-}0.49$	&
$\phantom{-}0.39$	&
$\phantom{-}0.30$	
\\
&
1.9 &
&
&
&
&
&
&
&
$\phantom{-}1.00$ &
$\phantom{-}0.83$	&
$\phantom{-}0.71$	&
$\phantom{-}0.61$	&
$\phantom{-}0.52$	
\\
&
2.0 &
&
&
&
&
&
&
&
&
$\phantom{-}1.00$ &
$\phantom{-}0.89$ &
$\phantom{-}0.80$ &
$\phantom{-}0.71$
\\
&
2.1
&
&
&
&
&
&
&
&
&
&
$\phantom{-}1.00$ &
$\phantom{-}0.96$	&
$\phantom{-}0.91$	
\\
&
2.2
&
&
&
&
&
&
&
&
&
&
&
$\phantom{-}1.00$ &
$\phantom{-}0.97$
\\
&
2.3
&
&
&
&
&
&
&
&
&
&
&
&
$\phantom{-}1.00$  
\\ \hline\hline
\end{tabular}
\end{center}
\end{table}

\section{Acknowlegdements}
We thank the KEKB group for the excellent operation of the
accelerator, the KEK cryogenics group for the efficient
operation of the solenoid, and the KEK computer group and
the National Institute of Informatics for valuable computing
and Super-SINET network support. We acknowledge support from
the Ministry of Education, Culture, Sports, Science, and
Technology of Japan and the Japan Society for the Promotion
of Science; the Australian Research Council and the
Australian Department of Education, Science and Training;
the National Science Foundation of China under contract
No.~10175071; the Department of Science and Technology of
India; the BK21 program of the Ministry of Education of
Korea and the CHEP SRC program of the Korea Science and
Engineering Foundation; the Polish State Committee for
Scientific Research under contract No.~2P03B 01324; the
Ministry of Science and Technology of the Russian
Federation; the Ministry of Higher Education, 
Science and Technology of the Republic of Slovenia;  
the Swiss National Science Foundation; the National Science Council and
the Ministry of Education of Taiwan; and the U.S.\
Department of Energy.

\end{document}